\documentclass[10pt,danish,english,journal]{IEEEtran}
\usepackage[T1]{fontenc}
\usepackage[latin9]{inputenc}
\usepackage{units}
\usepackage{amsthm}
\usepackage{amsmath}
\usepackage{amssymb}
\usepackage{esint}

\makeatletter

\providecommand{\tabularnewline}{\\}

\theoremstyle{plain}
\newtheorem{thm}{Theorem}
\theoremstyle{plain}
\newtheorem{lem}[thm]{Lemma}
\theoremstyle{plain}
\newtheorem{cor}[thm]{Corollary}
\theoremstyle{definition}
\newtheorem{example}[thm]{Example}
\theoremstyle{plain}
\newtheorem{conjecture}[thm]{Conjecture}
\theoremstyle{plain}
\newtheorem{prop}[thm]{Proposition}

\usepackage{amsfonts}
\usepackage{version}
\usepackage{pgf}
\usepackage{xcolor}
\usepackage{cite}
\usepackage{units}
\makeatother

\usepackage{babel}

\begin{document}

\title{Lower bounds on Information Divergence}

\author{Peter Harremoës and Christophe Vignat %
\thanks{P. Harremoës is with Copenhagen Business College, Denmark. C. Vignat
is with EPFL, Lausanne, Switzerland, and L.S.S., Supelec, France%
}}
\maketitle
\begin{abstract}
In this paper we establish lower bounds on information divergence
from a distribution to certain important classes of distributions
as Gaussian, exponential, Gamma, Poisson, geometric, and binomial.
These lower bounds are tight and for several convergence theorems
where a rate of convergence can be computed, this rate is determined
by the lower bounds proved in this paper. General techniques for getting
lower bounds in terms of moments are developed.
\end{abstract}

\section{Introduction and notations}

In 2004, O. Johnson and A. Barron have proved \cite{Johnson2004b}
that the rate of convergence in the information theoretic Central
Limit Theorem is upper bounded by $\nicefrac{c}{n}$ under suitable
conditions. P. Harremoës extended this work in \cite{Harremoes2005c}
based on a maximum entropy approach. Similar results have been obtained
for the convergence of binomial distributions to Poisson distributions.
Finally the rate of convergence of convolutions of distributions on
the unit circle toward the uniform distribution can be bounded. In
each of these cases lower bounds on information divergence in terms
of moments of orthogonal polynomials or trigonometric functions give
lower bounds on the rate of convergence. In this paper, we provide
more lower bounds on information divergence using mainly orthogonal
polynomials and the related exponential families.

We will identify $x!$ with $\Gamma\left(x+1\right)$ even when $x$
is not an integer. Similarly the generalized binomial coefficient
$\binom{x}{n}$ equals $x\left(x-1\right)\cdots\left(x-n+1\right)/n!$
when $x$ is not an integer. We use $\tau$ as short for $2\pi$.

\section{Moment calculations\label{SecMoment}}

Let $\left\{ Q_{\beta};\,\,\beta\in\Gamma\right\} $ denote an exponential
family of distributions such that the Radon-Nikodym derivative is
\[
\frac{\mathrm{d}Q_{\beta}}{\mathrm{d}Q_{0}}=\frac{\exp\left(\beta\cdot x\right)}{Z\left(\beta\right)}\]
 and where $\Gamma$ is the set of $\beta$ such that the \emph{partition
function} $Z$ is finite, i.e.\[
Z\left(\beta\right)=\int\exp\left(\beta\cdot x\right)~\mathrm{d}Q_{0}\left(x\right)<\infty.\]
The partition function $Z$ is also called the \emph{moment generating
function}. The parametrization $\beta\rightarrow Q_{\beta}$ is called
the natural parametrization. The mean value of the distribution $Q_{\beta}$
will be denoted $\mu_{\beta}.$ The distribution with mean value $\mu$
is denoted $Q^{\mu}$ so that $Q^{\mu_{\beta}}=Q_{\beta}.$ The inverse
of the function $\beta\rightarrow\mu_{\beta}$ is denoted $\hat{\beta}\left(\cdot\right)$
and equals the maximum likelihood estimate of the canonical parameter.
The variance of $x$ with respect to $Q^{\mu}$ is denoted $V\left(\mu\right)$
so that $\mu\rightarrow V\left(\mu\right)$ is the \emph{variance
function} of the exponential family. This variance function uniquely
characterizes the exponential family.

We note that $\beta\rightarrow\ln Z\left(\beta\right)$ is the \emph{cumulant
generating function} so that\begin{align*}
\frac{d}{d\beta}\ln Z\left(\beta\right)_{\mid\beta=0} & =\mathrm{\mathrm{E}}\left[X\right],\\
\frac{d^{2}}{d\beta^{2}}\ln Z\left(\beta\right)_{\mid\beta=0} & =\mathrm{Var}\left(X\right),\\
\frac{d^{3}}{d\beta^{3}}\ln Z\left(\beta\right)_{\mid\beta=0} & =\mathrm{E}\left[\left(X-\mathrm{E}\left[X\right]\right)^{3}\right].\end{align*}

\begin{lem}
\label{PeterG}Let $\left\{ Q_{\beta};\,\,\beta\in\Gamma\right\} $
denote an exponential family with \[
\frac{\mathrm{d}Q_{\beta}}{\mathrm{d}Q_{0}}=\frac{\exp\left(\beta\cdot x\right)}{Z\left(\beta\right)}.\]
 Then\end{lem}
\begin{enumerate}
\item for all $\mu$ and $\nu,$ \[
D\left(Q^{\mu}\Vert Q^{\nu}\right)=\frac{\left(\mu-\nu\right)^{2}}{2V\left(\eta\right)}\]
for some $\eta$ between $\mu$ and $\nu.$
\item for all $\alpha$ and $\beta\in\Gamma,$\[
D\left(Q_{\alpha}\Vert Q_{\beta}\right)=\frac{V\left(\mu_{\gamma}\right)}{2}\left(\alpha-\beta\right)^{2}\]
 for some $\gamma$ between $\alpha$ and $\beta,$ and\end{enumerate}
\begin{IEEEproof}
The two parts of the theorem are proved separately.
\begin{enumerate}
\item We consider the function \begin{align*}
g\left(t\right) & =D\left(Q^{t}\Vert Q^{\nu}\right)\\
 & =\left(\hat{\beta}\left(t\right)-\hat{\beta}\left(\nu\right)\right)\cdot t\\
 & \qquad+\ln Z\left(\hat{\beta}\left(\nu\right)\right)-\ln Z\left(\hat{\beta}\left(t\right)\right).\end{align*}
 The two first derivatives of this function are\begin{align*}
g^{\prime}\left(t\right) & =\frac{\mathrm{d}\hat{\beta}\left(t\right)}{\mathrm{d}t}t+\left(\hat{\beta}\left(t\right)-\hat{\beta}\left(\nu\right)\right)\\
 & \qquad-\frac{Z^{\prime}\left(\hat{\beta}\left(t\right)\right)}{Z\left(\hat{\beta}\left(t\right)\right)}\frac{\mathrm{d}\hat{\beta}\left(t\right)}{\mathrm{d}t}\\
 & =\hat{\beta}\left(t\right)-\hat{\beta}\left(\nu\right),\\
g^{\prime\prime}\left(t\right) & =\frac{1}{dt/d\hat{\beta}\left(t\right)}=\frac{1}{V\left(t\right)}.\end{align*}
 According to Taylor's formula there exists $\eta$ between $\mu$
and $\nu$ such that\begin{align*}
D & \left(Q^{\mu}\Vert Q^{\nu}\right)\\
 & =g\left(\nu\right)+\left(\mu-\nu\right)f^{\prime}\left(\nu\right)+\frac{1}{2}\left(\mu-\nu\right)^{2}f^{\prime\prime}\left(\eta\right)\\
 & =\frac{\left(\mu-\nu\right)^{2}}{2V\left(\eta\right)}.\end{align*}

\item The second part is proved in the same way as the first part.
\end{enumerate}
\end{IEEEproof}
\begin{cor}
Let $\beta\rightarrow Q_{\beta},\beta\in\Gamma$ denote an exponential
family with \[
\frac{\mathrm{d}Q_{\beta}}{\mathrm{d}Q_{0}}=\frac{\exp\left(\beta\cdot x\right)}{Z\left(\beta\right)}.\]
 If the variance function of the exponential family is increasing
then\[
D\left(Q^{\mu}\Vert Q^{\nu}\right)\geq\frac{\left(\mu-\nu\right)^{2}}{2V\left(\nu\right)}\]
for $\mu\leq\nu.$
\end{cor}
The binomial distributions, Poisson distributions, geometric distributions,
negative binomial distributions, inverse binomial distributions, and
generalized Poisson distributions are exponential families with at
most cubic variance functions \cite{Morris1982,Letac1990}. Using
the former corollary we can provide a lower bound on information divergence
in terms of the mean.
\begin{example}
The variance function of the Gaussian family is $V\left(\mu\right)=1$.
Hence, with $\Phi$ a standard normal random variable with probability
density $\frac{1}{\sqrt{\tau}}\exp\left(-\frac{x^{2}}{2}\right),$
\[
D\left(X\Vert\Phi\right)\geq\frac{1}{2}\mathrm{E}\left[X\right]^{2}\]
if $\mathrm{E}\left[X\right]\le0.$

This inequality actually holds if $X$ is Gaussian with variance 1;
using the exponential family based on the Gaussian distribution with
$x^{2}$ as sufficient statistics we get the inequality\[
D\left(X\Vert\Phi\right)\geq\frac{\left(Var\left(X\right)-1\right)^{2}}{6}\]
if $Var\left(X\right)\leq1.$ 
\end{example}
The next example is about the exponential distribution.
\begin{example}
\label{exa:The-Gamma-distribution}The \emph{Gamma distribution} with
shape parameter $\alpha$ and scale parameter $\beta$ reads\begin{equation}
\Gamma_{\alpha+1,\theta}\left(x\right)=\frac{\left(\frac{x}{\theta}\right)^{\alpha}}{\alpha!\theta}\exp\left(-\frac{x}{\theta}\right),\,\, x\ge0.\label{eq:gamma1}\end{equation}
The variance function of the Gamma distribution $V\left(m\right)=\frac{m^{2}}{\alpha+1}$
is increasing. Hence\[
D\left(X\Vert\Gamma_{\alpha+1,\theta}\right)\ge\frac{\left(\mathrm{E}\left[X\right]-m\right)^{2}}{\frac{2m^{2}}{\alpha+1}}\]
if $\mathrm{E}\left[X\right]\le m.$ Note that for $\alpha=0$ we
get the \emph{exponential distribution} as a special case.
\end{example}
The next example is about the binomial distribution.
\begin{example}
The \emph{binomial distribution} has point probabilities \[
bin\left(n,p,j\right)=\left(\begin{array}{c}
n\\
j\end{array}\right)p^{j}\left(1-p\right)^{n-j},\, j=0,1,2,\cdots,n.\]
The variance function is $V\left(m\right)=m-\nicefrac{m^{2}}{n}.$
The variance function has maximum for $m=\nicefrac{n}{2}.$ Hence\[
D\left(X\Vert bin\left(n,p,j\right)\right)\geq\frac{\left(\mathrm{E}\left[X\right]-np\right)^{2}}{2np\left(1-p\right)}\]
if \foreignlanguage{danish}{$\mathrm{E}\left[X\right]\leq np\leq\nicefrac{n}{2}$}
or if $\mathrm{E}\left[X\right]\geq np\geq\nicefrac{n}{2}$. For $p=\nicefrac{1}{2}$
the inequality\foreignlanguage{danish}{ \[
D\left(X\Vert bin\left(n,p,j\right)\right)\geq\frac{2\left(\mathrm{E}\left[X\right]-\frac{n}{2}\right)^{2}}{n}\]
}that holds for all random variables.
\end{example}
The next example is about the Poisson distribution.
\begin{example}
The \emph{Poisson distributions} with point probabilities\[
\frac{\lambda^{j}}{j!}\exp\left(-\lambda\right),\, j=0,1,2,\cdots\]
has variance function $V\left(\lambda\right)=\lambda$, which is increasing.
Hence\[
D\left(X\Vert Po\left(\lambda\right)\right)\geq\frac{\left(\mathrm{E}\left[X\right]-\lambda\right)^{2}}{2\lambda}\]
for $\mathrm{E}\left[X\right]\leq\lambda.$
\end{example}
\,
\begin{example}
The \emph{negative binomial distribution} $NB\left(r,p\right)$ with
success probability $p$ and number of failures $r$ has point probabilities\[
\binom{k+r-1}{k}\left(1-p\right)^{r}p^{k},\,\, k=0,1,2,\dots\]
Its variance function $V\left(m\right)=\frac{m\left(m+r\right)}{r}$
is increasing. Hence\[
D\left(X\Vert NB\left(r,p\right)\right)\ge\frac{\left(\mathrm{E}\left[X\right]-m\right)^{2}}{2\frac{m}{r}\left(m+r\right)}.\]
For $r=1$ we get the geometric distribution as a special case.
\end{example}
\,

The next examples involve cubic variance functions.
\begin{example}
The \emph{inverse Gaussian distribution} $IG$ has density \[
\left[\frac{\lambda}{\tau x^{3}}\right]^{1/2}\exp\left(\frac{-\lambda(x-\mu)^{2}}{2\mu^{2}x}\right).\]
The variance function $V\left(\mu\right)=\nicefrac{\mu^{3}}{\lambda}$
is increasing so \[
D\left(X\Vert IG\left(\mu,\lambda\right)\right)\geq\frac{\lambda\left(\mathrm{E}\left[X\right]-\mu\right)^{2}}{2\mu^{3}}\]
if \foreignlanguage{danish}{$\mathrm{E}\left[X\right]\leq\mu.$}
\end{example}
\,Similar results hold for the \emph{generalized Poisson distributions}
and for the \emph{inverse binomial distributions} \cite{Jain1971,Consul1973,Yanagimoto1989,Letac1990}.

\section{General results for Gamma distributions\label{SecLower}}

To simplify the exposition in this section we will assume that the
scale parameters $\theta$ of the Gamma distributions equal 1.

\subsection{A conjecture for the Gamma case}

The Gamma distribution reads\[
\Gamma_{\alpha+1,1}\left(x\right)=\frac{x^{\alpha}}{\alpha!}\exp\left(-x\right),\,\, x\ge0.\]
The Laguerre polynomials are given by the Rodrigues formula \[
L_{n}^{\alpha}\left(x\right)=\frac{x^{-\alpha}e^{x}}{n!}\frac{d^{n}}{dx^{n}}\left(x^{n+\alpha}e^{-x}\right),\,\alpha>-1.\]
The Laguerre polynomials are orthogonal with respect the Gamma distribution,
but they are not normalized and they do not all have positive leading
coefficient. We thus introduce the normalized Laguerre polynomials
by\[
\tilde{L}_{n}^{\alpha}\left(x\right)=\left(-1\right)^{n}\frac{L_{n}^{\alpha}\left(x\right)}{\binom{n+\alpha}{n}^{\nicefrac{1}{2}}}.\]
In Example \ref{exa:The-Gamma-distribution} we saw that inequality
\eqref{eq:gamma1} holds for any random variable $X$ satisfying $\mathrm{E}\left[\tilde{L}_{1}^{\alpha}\left(X\right)\right]<0.$
We conjecture that a similar result holds for the normalized Laguerre
polynomials of order 2.
\begin{conjecture}
\label{KonjMindre}For any random variable $X$ and for any $k\in\mathbb{N}$
we have \begin{equation}
D\left(X\Vert\Gamma_{\alpha+1,1}\right)\geq\frac{\mathrm{E}\left[\tilde{L}_{2}^{\alpha}\left(X\right)\right]^{2}}{2}\ \label{conjecture}\end{equation}
 if $\mathrm{E}\left[\tilde{L}_{2}^{\alpha}\left(X\right)\right]\leq0.$ \end{conjecture}
\begin{lem}
\label{LemmaNedreEpsilon}Let $\left\{ Q_{\beta};\,\,\beta\in\Gamma\right\} $
denote an exponential family with $x$ as sufficient statistics so
that \[
\frac{\mathrm{d}Q_{\beta}}{\mathrm{d}Q_{0}}=\frac{\exp\left(\beta\cdot x\right)}{Z\left(\beta\right)}.\]
 If $\mu_{0}=0$ and $V\left(0\right)=1$ and $\mathrm{E}_{Q_{0}}\left[X^{3}\right]>0$
then there exists $\varepsilon>0$ such that\[
D\left(Q^{\mu}\Vert Q_{0}\right)\geq\frac{\mu^{2}}{2}\]
 holds for $\mu\in\left[-\varepsilon,0\right].$ \end{lem}
\begin{IEEEproof}
From Lemma \ref{PeterG} we know that there exists $\eta$ between
$\mu$ and $\mu_{0}$ such that \begin{align*}
D\left(Q^{\mu}\Vert Q_{0}\right) & =D\left(Q^{\mu}\Vert Q^{\mu_{0}}\right)\\
 & =\frac{\left(\mu-\mu_{0}\right)^{2}}{2}\cdot\frac{1}{V\left(\eta\right)}.\end{align*}
 Therefore it is sufficient to prove there exists $\varepsilon>0$
such that $V\left(\eta\right)\leq1$ for $\left[-\varepsilon,0\right].$
This follows from the fact that \begin{align*}
\frac{\mathrm{d}V\left(\eta\right)}{\mathrm{d}\eta} & =\frac{\frac{\mathrm{d}V\left(\eta\right)}{\mathrm{d}\beta}}{\frac{\mathrm{d}\eta}{\mathrm{d}\beta}}\\
 & =\frac{\frac{\mathrm{d}^{3}}{\mathrm{d}\beta^{3}}\ln Z\left(\beta\right)}{\frac{\mathrm{d}^{2}}{\mathrm{d}\beta^{2}}\ln Z\left(\beta\right)}\\
 & =\frac{\mathrm{E}\left[\left(X-\eta\right)^{3}\right]}{\mathrm{Var}\left(X\right)}\end{align*}
 where the mean and variance are taken with respect to the element
in the exponential family with mean $\eta.$ Since $\frac{\mathrm{\mathrm{E}}\left[X^{3}\right]}{\mathrm{Var}\left(X\right)}>0$
for $\beta=0$ we have that $\frac{\mathrm{E}\left[\left(X-\eta\right)^{3}\right]}{\mathrm{Var}\left(X\right)}>0$
for $\beta$ in a neighborhood of 0 so $V\left(\eta\right)$ is an
increasing of $\eta$ and the result then follows from $V\left(0\right)=1.$ 
\end{IEEEproof}
We can now formulate the following result.
\begin{prop}
For all $n\in\mathbb{N}_{0}$ and all $\alpha>-1$ we have \[
\int_{0}^{\infty}\left(\tilde{L}_{n}^{\alpha}\left(x\right)\right)^{3}\Gamma_{\alpha+1,1}\left(x\right)\, dx>0.\]
\end{prop}
\begin{IEEEproof}
We have\[
\int_{0}^{\infty}\left(\tilde{L}_{n}^{\alpha}\left(x\right)\right)^{3}\Gamma_{\alpha}\left(x\right)dx=\int_{0}^{\infty}\left(-1\right)^{3n}\frac{\left(L_{n}^{\alpha}\left(x\right)\right)^{3}}{\binom{n+\alpha}{n}^{\nicefrac{3}{2}}}\Gamma_{\alpha+1,1}\left(x\right)dx\]
\[
=\left(-1\right)^{n}\binom{n+\alpha}{n}^{-\nicefrac{3}{2}}\int_{0}^{\infty}\left(L_{n}^{\alpha}\left(x\right)\right)^{3}\Gamma_{\alpha+1,1}\left(x\right)\, dx,\]
which, according to \cite[p. 57]{Askey1977}, is strictly positive.\end{IEEEproof}
\begin{thm}
For any $n\in\mathbb{N}_{0}$ and any $\alpha>-1$ there exists $\epsilon>0$
that may depend on $\alpha$ and $n$ such that\[
D\left(X\Vert\Gamma_{\alpha+1,1}\left(X\right)\right)\geq\frac{1}{2}\left(\mathrm{E}\left[\tilde{L}_{n}^{\alpha}\left(x\right)\right]\right)^{2}\]
for any random variable $X$ satisfying $\mathrm{E}\left[\tilde{L}_{n}^{\alpha}\left(x\right)\right]\in\left[-\epsilon,0\right].$
\end{thm}
In the Gaussian case, we have the similar
\begin{cor}
For any $n\in\mathbb{N}_{0}$ there exists $\epsilon>0$ such that
\[
D\left(X\Vert\Phi\right)\geq\frac{1}{2}\left(\mathrm{E}\left[H_{2n}\left(X\right)\right]\right)^{2}\]
for any random variable $X$ satisfying $H_{2n}\left(X\right)\in\left[-\epsilon,0\right].$
\end{cor}
This inequality has previously been proved by considering the Hermite
polynomials as limits of Poisson-Charlier polynomials for which a
similar inequality holds \cite{Harremoes2008c}.

\section{Laguerre polynomials of degree 2}

We shall use the following lemma.
\begin{lem}
\label{lem:checkint}Assume that \begin{equation}
\int_{0}^{\infty}\left(\tilde{L}_{k}^{\alpha}\left(X\right)\right)^{2}\exp\left(\beta_{0}\tilde{L}_{k}^{\alpha}\left(X\right)\right)\frac{x^{\alpha}}{\alpha!}\exp\left(-x\right)\, dx\leq1\label{eq:check}\end{equation}

\end{lem}
then the conjecture holds for all $\mathrm{E}\left[\tilde{L}_{k}^{\alpha}\left(X\right)\right]\in\left[\beta_{0},0\right].$
\begin{IEEEproof}
Let $Q_{\beta}$ denote the distribution with density\[
\frac{\mathrm{d}Q_{\beta}}{\mathrm{d}\Gamma_{\alpha+1,1}}=\frac{\exp\left(\beta\cdot\tilde{L}_{k}^{\alpha}\left(x\right)\right)}{Z\left(\beta\right)}.\]
We have to prove that \[
D\left(Q_{\beta}\Vert\Gamma_{\alpha+1,1}\right)\geq\frac{1}{2}\left(\mu_{\beta}\right)^{2}.\]
for $\mu_{\beta}\in\left[\beta_{0},0\right]$. We have $D\left(Q_{\beta}\Vert\Gamma_{\alpha+1,1}\right)=\beta\mu_{\beta}-\ln\left(Z\left(\beta\right)\right)$
and $\mu_{\beta}=\frac{Z'\left(\beta\right)}{Z\left(\beta\right)}.$
The inequality is satisfied for $\beta=0$ so we differentiate with
respect to $\beta$ and have to prove that \[
\mu_{\beta}+\beta\frac{d\mu_{\beta}}{d\beta}-\frac{Z'\left(\beta\right)}{Z\left(\beta\right)}\leq\frac{1}{2}\cdot2\mu_{\beta}\frac{d\mu_{\beta}}{d\beta}\]
which is equivalent to\[
\beta\leq\mu_{\beta}.\]
Since we have assumed that $\mu_{\beta}\in\left[\beta_{0},0\right]$
it is sufficient to prove the inequality for $\beta\in\left[\beta_{0},0\right].$
The inequality is satisfied for $\beta=0$ so we differentiate once
more so that we have to prove the inequality\foreignlanguage{danish}{\[
1\geq d\mu_{\beta}=\frac{Z^{\prime\prime}\left(\beta\right)Z\left(\beta\right)-\left(Z^{\prime}\left(\beta\right)\right)^{2}}{\left(Z\left(\beta\right)\right)^{2}}.\]
}

Hence it is sufficient to prove that \[
\frac{Z^{\prime\prime}\left(\beta\right)}{Z\left(\beta\right)}\leq1\]
which is equivalent to \[
Z''\left(\beta\right)\leq Z\left(\beta\right).\]
 Since $Z\left(\beta\right)\geq1$ for all $\beta$ so it is sufficient
to prove that $Z''\left(\beta\right)\leq1$ for $\beta\in\left[\beta_{0},0\right]$.
The function $\beta\rightarrow Z''\left(\beta\right)$ is convex and
\[
Z''\left(0\right)=\int_{0}^{\infty}\left(\tilde{L}_{k}^{\alpha}\left(X\right)\right)^{2}\frac{x^{\alpha}}{\alpha!}\exp\left(-x\right)\, dx=1.\]
Therefore it is sufficient to check that $Z''\left(\beta_{0}\right)\leq1$,
which is exactly what is stated in (\ref{eq:check}).
\end{IEEEproof}

\subsection{Large values of the shape parameter}

If the scale parameter is fixed at 1 and the shape parameter tends
to infinity then the Gamma distribution will tend to a Gaussian. We
know that one can get a lower bound on the information divergence
in terms of the Hermite polynomial of order 2 so we should expect
this also to hold for large values of the shape parameter. This is
indeed the case as stated i the following theorem.
\begin{thm}
For any $\alpha\geq6\nicefrac{1}{2}$ \[
D\left(X\Vert\Gamma_{\alpha+1,1}\left(X\right)\right)\geq\frac{1}{2}\left(\mathrm{E}\left[\tilde{L}_{2}^{\alpha}\left(X\right)\right]\right)^{2}\]
for any random variable $X$ satisfying $\mathrm{E}\left[\tilde{L}_{n}^{\alpha}\left(x\right)\right]\leq0.$\end{thm}
\begin{IEEEproof}
Let $\beta_{0}$ denote the negative solution to the equation $\beta^{2}\exp\left(\beta^{2}\right)=1$.
The value is approximately $\beta_{0}=-0.75309$~. The function $f\left(x\right)=x^{2}\exp\left(\beta_{0}x\right)$
is decreasing for $x\in\left]-\infty,0\right]$, increasing for $x\in\left[0,-\nicefrac{2}{\beta_{0}}\right]$and
decreasing for $x\in\left[-\nicefrac{2}{\beta_{0}},\infty\right[$.
The local maximum in $x=-\nicefrac{2}{\beta_{0}}$ has the value $0.9545<1.$
According to the definition of $\beta_{0}$ we have $f\left(\beta_{0}\right)=1$
so f $f\left(x\right)\leq1$ for $x\geq\beta_{0}.$ The second normalized
Laguerre polynomial is\begin{align*}
\tilde{L}_{n}^{\alpha}\left(x\right) & =\frac{L_{n}^{\alpha}\left(x\right)}{\left(\frac{\left(2+\alpha\right)\left(1+\alpha\right)}{2}\right)^{\nicefrac{1}{2}}}\\
 & =\frac{x^{2}-2\left(\alpha+2\right)x+\left(\alpha+2\right)\left(\alpha+1\right)}{\left(2\left(2+\alpha\right)\left(1+\alpha\right)\right)^{\nicefrac{1}{2}}}.\end{align*}
The minimum is attained for $x=2\left(\alpha+1\right)$ and has the
value \[
-2^{-\nicefrac{1}{2}}\left(1+\frac{1}{\alpha+1}\right)^{\nicefrac{1}{2}}.\]
This is an increasing function of $\alpha$ that tends to $-2^{-\nicefrac{1}{2}}>\beta_{0}$
for $x$ tending to $\infty.$ We solve the equation\[
-2^{-\nicefrac{1}{2}}\left(1+\frac{1}{\alpha+1}\right)^{\nicefrac{1}{2}}=\beta_{0}\]
and get\[
\alpha_{0}=\frac{1}{2\beta_{0}^{2}-1}-1=6.4466\,.\]
Therefore \foreignlanguage{danish}{$\tilde{L}_{n}^{\alpha}\left(x\right)\geq\beta_{0}$
for all $x$ if $\alpha\geq\alpha_{0}.$ Hence $f\left(\tilde{L}_{n}^{\alpha}\left(x\right)\right)\leq1$
for all $x$ if $\alpha\geq\alpha_{0}.$ In particular\begin{multline*}
\int_{0}^{\infty}\left(\tilde{L}_{2}^{\alpha}\left(x\right)\right)^{2}\exp\left(\beta_{0}\tilde{L}_{k}^{\alpha}\left(x\right)\right)\frac{x^{\alpha}}{\alpha!}\exp\left(-x\right)\, dx\\
=\int_{0}^{\infty}f\left(\tilde{L}_{2}^{\alpha}\left(x\right)\right)\frac{x^{\alpha}}{\alpha!}\exp\left(-x\right)\, dx\leq1.\end{multline*}
This proves that the inequality holds whenever $\mathrm{E}\left[\tilde{L}_{2}^{\alpha}\left(X\right)\right]\in\left[\beta_{0},0\right].$
The condition $\mathrm{E}\left[\tilde{L}_{2}^{\alpha}\left(X\right)\right]\geq\beta_{0}$
is automatically fulfilled if $\alpha\geq\alpha_{0}$ and the theorem
follows.}
\end{IEEEproof}

\subsection{Chi square distributions}

\selectlanguage{danish}%
The $\chi^{2}$-distributions are Gamma distributions with half integral
value of $\alpha$ and scale parameter $\nicefrac{1}{2}.$ We will
check our conjecture for $\alpha<6\nicefrac{1}{2}$ and half integral
values. For notational convenience we will assume that the shape parameter
is 1 and note that results for $\chi^{2}$-distributions are obtained
by a simple scaling. According to Lemma \ref{lem:checkint} it is
sufficient to calculate the integral (\ref{eq:check}) when \[
\beta_{0}=\min_{x}\tilde{L}_{2}^{\alpha}\left(X\right)=-2^{-\nicefrac{1}{2}}\left(1+\frac{1}{\alpha+1}\right)^{\nicefrac{1}{2}}.\]
The results are given in the following.

\begin{tabular}{|c|c|c|}
\hline 
\selectlanguage{english}%
$\alpha$\selectlanguage{danish}
 & \selectlanguage{english}%
$\beta_{0}$\selectlanguage{danish}
 & \selectlanguage{english}%
$\int$\selectlanguage{danish}
\tabularnewline
\hline
\hline 
\selectlanguage{english}%
$-\nicefrac{1}{2}$\selectlanguage{danish}
 & \selectlanguage{english}%
-1.225\selectlanguage{danish}
 & \selectlanguage{english}%
0.95407\selectlanguage{danish}
\tabularnewline
\hline 
\selectlanguage{english}%
0\selectlanguage{danish}
 & \selectlanguage{english}%
-1\selectlanguage{danish}
 & \selectlanguage{english}%
0.63113\selectlanguage{danish}
\tabularnewline
\hline 
\selectlanguage{english}%
$\nicefrac{1}{2}$\selectlanguage{danish}
 & \selectlanguage{english}%
-0.9129\selectlanguage{danish}
 & \selectlanguage{english}%
0.55406\selectlanguage{danish}
\tabularnewline
\hline 
\selectlanguage{english}%
1\selectlanguage{danish}
 & \selectlanguage{english}%
-0.8660\selectlanguage{danish}
 & \selectlanguage{english}%
0.52046\selectlanguage{danish}
\tabularnewline
\hline 
\selectlanguage{english}%
$1\nicefrac{1}{2}$\selectlanguage{danish}
 & \selectlanguage{english}%
-0.8367\selectlanguage{danish}
 & \selectlanguage{english}%
0.5018\selectlanguage{danish}
\tabularnewline
\hline 
\selectlanguage{english}%
2\selectlanguage{danish}
 & \selectlanguage{english}%
-0.8165\selectlanguage{danish}
 & \selectlanguage{english}%
0.48997\selectlanguage{danish}
\tabularnewline
\hline 
\selectlanguage{english}%
$2\nicefrac{1}{2}$\selectlanguage{danish}
 & \selectlanguage{english}%
-0.8018\selectlanguage{danish}
 & \selectlanguage{english}%
0.48181\selectlanguage{danish}
\tabularnewline
\hline 
\selectlanguage{english}%
3\selectlanguage{danish}
 & \selectlanguage{english}%
-0.7906\selectlanguage{danish}
 & \selectlanguage{english}%
0.47584\selectlanguage{danish}
\tabularnewline
\hline 
\selectlanguage{english}%
$3\nicefrac{1}{2}$\selectlanguage{danish}
 & \selectlanguage{english}%
-0.7817\selectlanguage{danish}
 & \selectlanguage{english}%
0.47128\selectlanguage{danish}
\tabularnewline
\hline 
\selectlanguage{english}%
4\selectlanguage{danish}
 & \selectlanguage{english}%
-0.7746\selectlanguage{danish}
 & \selectlanguage{english}%
0.46769\selectlanguage{danish}
\tabularnewline
\hline 
\selectlanguage{english}%
$4\nicefrac{1}{2}$\selectlanguage{danish}
 & \selectlanguage{english}%
-0.7687\selectlanguage{danish}
 & \selectlanguage{english}%
0.46478\selectlanguage{danish}
\tabularnewline
\hline 
\selectlanguage{english}%
5\selectlanguage{danish}
 & \selectlanguage{english}%
-0.7638\selectlanguage{danish}
 & \selectlanguage{english}%
0.46238\selectlanguage{danish}
\tabularnewline
\hline 
\selectlanguage{english}%
$5\nicefrac{1}{2}$\selectlanguage{danish}
 & \selectlanguage{english}%
-0.7596\selectlanguage{danish}
 & \selectlanguage{english}%
0.46037\selectlanguage{danish}
\tabularnewline
\hline 
\selectlanguage{english}%
6\selectlanguage{danish}
 & \selectlanguage{english}%
-0.7559\selectlanguage{danish}
 & \selectlanguage{english}%
0.45865\selectlanguage{danish}
\tabularnewline
\hline
\end{tabular}

As we see all values of the integral are less than 1so the conjecture
holds for all half integral values of $\alpha.$ \foreignlanguage{english}{This
gives us the following theorem.}
\selectlanguage{english}%
\begin{thm}
Assume that $\alpha>-1$ and that $2\alpha$ is an integer. Then,
for any random variable satisfying $\mathrm{E}\left[\tilde{L}_{2}^{\alpha}\left(X\right)\right]\leq0,$
we have \[
D\left(X\Vert\Gamma_{\alpha+1,1}\right)\geq\frac{1}{2}\left(\mathrm{E}\left[\tilde{L}_{2}^{\alpha}\left(X\right)\right]\right)^{2}.\]
\end{thm}
\begin{example}
For $\alpha=0$ we get the exponential distribution with density\[
\exp\left(-x\right),\, x>0.\]
The Laguerre polynomial of order two is $L_{2}\left(x\right)=\frac{1}{2}x^{2}-2x+1.$
We will rewrite our inequality in terms of mean and variance. For
any random variable satisfying $Var\left(X\right)\leq1$ and $E\left[X\right]=1$
we get the inequality \[
D\left(X\Vert Exp\left(1\right)\right)\geq\frac{1}{8}\left(\mathrm{Var\left(X\right)-1}\right)^{2}.\]

\end{example}
The $\chi^{2}$-distribution with 1 degree of freedom corresponds
to a Gamma distribution with shape parameter $\alpha+1=\nicefrac{1}{2}$
and scale parameter $2.$ It has density\[
\frac{x^{-\nicefrac{1}{2}}}{\tau^{1/2}}\exp\left(-\frac{x}{2}\right)\]
This distribution is important because it is the distribution of the
square of a standard Gaussian random variable. Hence, results for
the $\chi^{2}$ distribution translate into results for Hermite moments.
In order to follow the notation from the previous section we first
prove results for the Gamma distribution with shape parameter $\alpha+1=\nicefrac{1}{2}$
and scale parameter $1$ and then translate the results. 

We have \[
L_{2}^{-\nicefrac{1}{2}}\left(x\right)=\frac{x^{2}}{2}-\frac{3}{2}x+\frac{3}{8}\]
and the normalized version \[
\tilde{L}_{2}^{-\nicefrac{1}{2}}\left(x\right)=\frac{x^{2}-3x+\frac{3}{4}}{\left(\nicefrac{3}{2}\right)^{\nicefrac{1}{2}}}.\]
This gives us the following theorem.
\begin{thm}
For any random variable satisfying $\mathrm{E}\left[\tilde{L}_{2}^{-1/2}\left(X\right)\right]\leq0,$
we have \[
D\left(X\Vert\Gamma_{1/2,1}\right)\geq\frac{1}{2}\left(\mathrm{E}\left[\tilde{L}_{2}^{-1/2}\left(X\right)\right]\right)^{2}.\]
\end{thm}
\begin{cor}
For a random variable $X$ satisfying $\mathrm{E}\left[X\right]=1$
and $\mathrm{Var}\left(X\right)\leq2$ we have\[
D\left(X\Vert\chi_{1}^{2}\right)\geq\frac{\left(\mathrm{Var}\left(X\right)-2\right)^{2}}{48}.\]
\end{cor}
\begin{IEEEproof}
The result follows from the following computation.\begin{align*}
D\left(X\Vert\chi_{1}^{2}\right) & =D\left(\frac{X}{2}\Vert\Gamma_{\alpha+1,1}\right)\\
 & \geq\frac{1}{2}\left(\mathrm{E}\left[\tilde{L}_{2}^{-\nicefrac{1}{2}}\left(\frac{X}{2}\right)\right]\right)^{2}\\
 & =\frac{1}{2}\left(\mathrm{E}\left[\frac{\left(\frac{X}{2}\right)^{2}-3\left(\frac{X}{2}\right)+\frac{3}{4}}{\left(\frac{3}{2}\right)^{1/2}}\right]\right)^{2}\\
 & \mathrm{=\frac{1}{48}}\left(\mathrm{Var}\left(X\right)+\mathrm{E}\left[x\right]^{2}-6\mathrm{E}\left[X\right]+3\right)^{2}\\
 & =\frac{\left(\mathrm{Var}\left(X\right)-2\right)^{2}}{48}\,.\end{align*}

\end{IEEEproof}
These inequalities can be translated into inequalities for Hermite
polynomials.
\begin{cor}
For any random variable satisfying $\mathrm{E}\left[H_{4}\left(X\right)\right]\leq0$
we have \[
D\left(X\Vert\Phi\right)\geq\frac{1}{2}\left(\mathrm{E}\left[H_{4}\left(X\right)\right]\right)^{2}.\]

If $Var\left(X\right)=1$ this is equivalent to\begin{equation}
D\left(X\Vert\Phi\right)\geq\frac{\kappa^{2}}{48}\label{eq:platykurtic}\end{equation}
if $X$ is platykurtic and $\kappa$ denotes the excess kurtosis.
\end{cor}
The inequality (\ref{eq:platykurtic}) was proved in \cite{Harremoes2005c}
with a different technique.

\section{Counterexample}

With all these positive results in mind one may conjecture that \begin{equation}
D\left(X\Vert\Gamma_{\alpha+1,1}\right)\geq\frac{\mathrm{E}\left[\tilde{L}_{k}^{\alpha}\left(X\right)\right]^{2}}{2}\ \label{conjecture-1}\end{equation}
 would hold for all $k$ as long as $\mathrm{E}\left[\tilde{L}_{k}^{\alpha}\left(X\right)\right]\leq0$,
but this is not the case. Here we will describe a counterexample for
$k=3$ and $\alpha=-1/2$. We will fix \foreignlanguage{danish}{$\mathrm{E}\left[\tilde{L}_{k}^{\alpha}\left(X\right)\right]=-3.$
In this case the information projection of $\Gamma_{\nicefrac{1}{2},1}$
onto the set of distributions satisfying $\mathrm{E}\left[\tilde{L}_{k}^{\alpha}\left(X\right)\right]=-3$
equals the distribution $Q_{\beta}$ with density\[
\frac{\mathrm{d}Q_{\beta}}{\mathrm{d}\Gamma_{\nicefrac{1}{2},1}}\left(x\right)=\frac{\exp\left(\beta\tilde{L}_{3}^{\nicefrac{1}{2}}\left(x\right)\right)}{\int_{0}^{\infty}\exp\left(\beta\tilde{L}_{3}^{\nicefrac{1}{2}}\left(x\right)\Gamma_{\alpha+1,\theta}\left(x\right)\,\mathrm{d}x\right)}\,.\]
Numerical calculations gives $\beta=-1.83125$ and $D\left(Q_{\beta}\Vert\Gamma_{\nicefrac{1}{2},1}\right)=3.3195$,
which is not greater than $\frac{1}{2}\left(-3\right)^{2}.$ The counterexample
implies that there exists a random variable $X$ such that \[
D\left(X\Vert\Phi\right)\not\geq\frac{1}{2}\left(\mathrm{E}\left[H_{6}\left(X\right)\right]\right)^{2}.\]
}

\section{Acknowledgment}

Lemma \ref{PeterG} was developed in close collaboration with Peter
Grünwald. We thank Oliver Johnson and Ioannis Kontoyiannis for useful
discussions.

\bibliographystyle{ieeetr}
\bibliography{S:/Servage.net/harremoes.dk/www/Peter/bibtex/database1}

\begin{thebibliography}{1}

\bibitem{Johnson2004b}
O.~Johnson and A.~R. Barron, ``Fisher information inequalities and the central
  limit theorem,'' {\em Probability Theory and Related Fields}, vol.~129,
  pp.~391--409, April 2004.

\bibitem{Harremoes2005c}
P.~Harremo{\"e}s, ``Lower bounds on divergence in central limit theorem,'' {\em
  Electronic Notes in Discrete Mathematics}, vol.~21, pp.~309--313, Aug. 2005.

\bibitem{Morris1982}
C.~Morris, ``Natural exponential families with quadratic variance functions,''
  {\em Ann. Statist.}, vol.~10, pp.~65--80, 1982.

\bibitem{Letac1990}
G.~Letac and M.~Mora, ``Natural real exponential families with cubic variance
  functions,'' {\em Ann. Stat.}, vol.~18, no.~1, pp.~1--37, 1990.

\bibitem{Jain1971}
G.~C. Jain and P.~C. Consul, ``A generalized negative binomial distribution,''
  {\em SIAM Journal on Applied Mathematics}, vol.~21, no.~4, pp.~501--513,
  1971.

\bibitem{Consul1973}
P.~C. Consul and G.~C. Jain, ``A generalization of the {P}oisson
  distribution,'' {\em Technometrics}, vol.~15, pp.~791--799, 1973.

\bibitem{Yanagimoto1989}
T.~Yanagimoto, ``The inverse binomial distribution as a statistical model,''
  {\em Communications in Statistics: Theory and Methods}, vol.~18,
  pp.~3625--3633, 1989.

\bibitem{Askey1977}
R.~Askey and G.~Gasper, ``Convolution structure for {L}aguerre polynomials,''
  {\em Journal D'Analyse Math\'ematique}, vol.~31, pp.~48--68, 1977.

\bibitem{Harremoes2008c}
P.~Harremo{\"e}s, O.~Johnson, and I.~Kontoyiannis, ``Thinning and information
  projection,'' in {\em International Symposium on Information Theory},
  pp.~2644--2648, IEEE, July 2008.

\end{thebibliography}

\end{document}